\documentstyle [a4,12pt]{article}
\begin{document}
\begin{center}
{\large \bf ABOUT SPONTANEOUS SYMMETRY
BREAKING IN THE FIELD THEORY WITH FUNDAMENTAL MASS }\\
 \vspace{0,5cm}
 {\bf Umida  Khodjaeva}\\
 \vspace{0,5cm}
 Department of Theoretical Physics and
Computer Sciences\\ of Samarkand State University.\\
15,
University blvd, 703004 Samarkand, Uzbekistan.\\
 umida@samdu.uz
\end{center}

{\bf Abstract.} The simple examples of spontaneous breaking of
various symmetries for the scalar theory with fundamental mass
have been considered. Higgs generalizations on "fundamental mass"
were introduced into the theory on a basis of the five-dimensional
de Sitter
space.\\

The concept of mass having its root in great antiquity still
remains fundamental. Every theoretical and experimental research
in classical physics and quantum physics, related to mass is a
step to insight of the nature. Besides mass, other fundamental
constants, such as Planck's  constant $\hbar$  and speed of light
$c$ , also play the most important role in modern theories. The
first one is related to quantum mechanics, and the second one is
related to the theory of relativity.

Characteristics and interactions of elementary particles (EP) can
be described more or less in terms of local fields (LF) which in
their turn regard to low representations of corresponding compact
groups of symmetry. Concept of LF essentially is a synonym of
concept EP. At present elementary particles are such kind of
particles (real and hypothetical), characteristics and
interactions of which could be adequately described in terms of
LF. As we know, mass of EP  $m$ is Kazimir's operator of
noncompact Poincare group, and those representations of the given
group which are used in the quantum field theory (QFT), can take
any values in an interval $0 \leq m < \infty $. Nowadays two
particles mentioned as EP can have masses different from each
other on many orders. Formally standard QFT remains logically
irreproachable circuit in cases when masses of particles can be
comparable to masses of macro matters. Modern QFT does not forbid
such physically nonsensical extrapolation. Probably it is the
basic defect of the theory?

In 1965 M.A.Markov put forward a hypothesis \cite{1} according to
which the spectrum of masses of EP should break on "planck mass"

\begin{equation}
m<m_{Planck}=\sqrt{\frac{\hbar c}{G}}\approx 10^{19} GeV,
\end{equation}
here $\hbar$ , $c$   are known universal constants and $G$ is a
gravitational constant. The particles of limiting mass
$m=m_{Planck}$ named by M.A.Markov as "maximons" are called to
play a special role in the world of elementary particles. The
concept of "maximon" is assumed as a basis of Markov's script of
the early universe \cite{2}. It is significant that in relation to
QFT Markov's restriction (1) acts as an additional
phenomenological condition. It does not affect structure of this
theory in any way, and even for the description of maximon the
standard theoretical-field device is used. New version of QFT, in
basis of which the postulate-M.A.Markov's principle about
limitation of mass of elementary particles (1) is put alongside
with traditional quantum and relativistic postulates, has been
worked out by V.G.Kadyshevsky \cite{3}. The key role in the
approach developed by him belongs to 5-dimensional configuration
representation. Remaining inherently four-dimensional, the theory
assumes the original local lagrange formulation in which
dependence of fields on auxiliary fifth coordinate also is found
as local. Internal symmetries in this formalism generate the gauge
transformations localized in the same 5-dimensional configuration
space. Thus Markov's condition is written down as $m\leq M$ ,
considering limiting mass $M$ simply as  a certain new universal
constant of the theory, so-called "fundamental mass" (FM). EP with
$m=M$ are still called as maximons. In the limit
$M\rightarrow\infty$ new QFT coincides with the usual field theory
in which the spectrum of particles is unlimited. On a strict
mathematical basis new parameter FM is entered in QFT which.
Together with parameters of the standard quantum theory this
parameter will play an essential role in high energy physics
\cite{4}. In work \cite{5} geometrical interpretation of effect of
spontaneous breaking of symmetry which plays a key role in
standard model is advanced. This approach is related to an
effective utilization in device QFT of 4-pulse de Sitter and
anti-de Sitter's spaces with constant curvature. In our works
\cite{6} simple examples of spontaneous breakings of various
symmetries for the scalar theory with FM have been considered and
Higgs generalizations on FM are cited.

In the given work we shall continue research on the basis of
simple examples of spontaneous breakings of various symmetries for
the scalar theory with FM. For this purpose we use lagrange
formalism from works \cite{3,4}.

Formulation of QFT with FM, discussed in work \cite{4}, is based
on the quantum version of the de Sitter's equation, that is on the
5-dimensional equation of a field:
\begin{equation}
\left[\frac{\partial^2}{\partial x^{\mu}\partial x_{\mu}}-
\frac{\partial^2}{\partial x^2_5}-\frac{M^2c^2}{\hbar^2}\right]
\Phi(x,x^5)=0
\end{equation}
$$
\small{\qquad\qquad\qquad \mu =0,1,2,3.}
$$
To every field in 5-space a wave function  $\Phi(x,x^5)$
submitting with the equation (2) is compared. This is equivalent
to the statement that $\Phi(x,x^5)$ the field in usual space-time
is described by wave function with the double number of
components:
\begin{equation}\Phi(x,x^5)\leftrightarrow \left(
\begin{array}{c}
\Phi(x,0) \\
\partial \Phi(x,0)/\partial x^{5} .
\end{array}
\right)\equiv\left(\begin{array}{c}
\Phi(x) \\
\chi(x)
\end{array}
\right).
\end{equation}
Typical for this circuit the doubling of number of field degrees
of freedom disappears at $M\rightarrow\infty$. At finite $M$ the
analogue of a usual field variable should be considered
$\Phi(x)=\Phi(x,0)$, $\chi(x)= \partial \Phi(x,0)/\partial x^{5}$
and function  is auxiliary.

Now we shall consider simple examples of spontaneous breakings of
various symmetries for the scalar theory with FM. The lagrangian
of the real scalar field in frameworks of QFT with FM has the form
\cite{4}:
 \begin{equation}
  L(x,M)=\frac{1}{2} [[ \frac{\partial \Phi (x)}{\partial
  x_{n}}]^{2}+m^{2}\Phi^{2}(x)+M^{2}[\chi
  (x)-\cos\mu\Phi(x)]^{2}].
\end{equation}
Taking into account interaction in (3), we can (4) write
following:
 \begin{equation}
  L(x,M)=\frac{1}{2} [ \frac{\partial \varphi(x)}{\partial
  x_{\mu}}]^{2}- \frac{1}{2}m^{2}\varphi^{2}(x)-\frac{1}{2}M^{2}(\chi
  (x)-\cos\mu\varphi(x))^{2}-\chi(x)U(\varphi(x),
\end{equation}
here $\cos\mu=\surd\overline{1-\frac{m^{2}}{M^{2}}}$ , where $m$ -
mass of the particles, described by field $\varphi(x)$, and
$\chi(x)$  is the auxiliary field, playing a role in interaction,
$M$- fundamental mass and  $U(\varphi(x))$ is the unknown function
describing interactions of particles.

Whether is possible to choose the interaction
$L_{int}=\chi(x)U(\varphi(x))$ between fields $\varphi(x)$ and
$\chi(x)$ that Higgs potential for a field $\varphi(x)$  exists at
exception of a field $\chi(x)$? Free lagrangian (5) is invariant
under transformation $\varphi(x)\rightarrow -\varphi(x)$ and
$\chi(x)\rightarrow -\chi(x)$. But thus is necessary to demand,
that $U(-\varphi)\rightarrow -U(\varphi)$, that $U(\varphi)$  is
an odd function of $\varphi(x)$. Action for (5) is possible to be
written as:
  \begin{equation}
  S(x,M)=\int{\{\frac{1}{2} [ \frac{\partial \varphi(x)}{\partial
  x_{\mu}}]^{2}- \frac{1}{2}m^{2}\varphi^{2}(x)-\frac{1}{2}M^{2}(\chi
  (x)-\cos\mu\varphi(x))^{2}+\chi(x)U(\varphi(x)\}}d\varphi(x).
\end{equation}
If we differentiate (6) with respect to $\chi(x)$ , we find:
 \begin{equation}
\chi(x)=\cos\mu\varphi(x)+\frac{U(\varphi(x))}{M^{2}}.
\end{equation}
Substituting (7) in (6), we have:
 \begin{equation}
  L_{tot}(\varphi(x))=\frac{1}{2} [ (\frac{\partial \varphi(x)}{\partial
  x_{\mu}})^{2}- m^{2}\varphi^{2}(x)+\frac{U^{2}(\varphi(x))}{M^{2}}
  +2U(\varphi(x))\cos\mu\varphi(x)],
  \end{equation}
that is invariant to $\varphi(x)\rightarrow -\varphi(x)$.

From breakings of discrete symmetry for usual scalar field it is
known that Higgs potential looks like:
\begin{equation}
V(\varphi(x))=-\frac{1}{2}m^{2}\varphi^{2}(x)+\frac{1}{4}{\lambda^{2}\varphi^{4}(x)},
\end{equation}
where $\lambda $ is the dimensionless constant describing
interaction between particles.

Let find a kind of $U(\varphi(x))$ function that in (8) Higgs
potential to appear. We shall consider the lagrangian (8) at
$m\rightarrow im$, then $
\cos\mu=\sqrt{1-\frac{m^{2}}{M^{2}}}\rightarrow ch\mu^{'}=
\sqrt{1+\frac{m^{2}}{M^{2}}} $. Potential energy (9) shall look
like:
\begin{equation}
V(\varphi(x))=\frac{1}{2}m^{2}\varphi(x)^{2}-\frac{U^{2}(\varphi(x))}{2
M^{2}}-U(\varphi(x))ch\mu^{'}\varphi(x)
\end{equation}
Comparing (10) and (9), for $U(\varphi(x))$ we have two different
roots (real and imaginary) at  $\varphi^{2}<\frac{2
M^{2}ch^{2}\mu^{'}}{\lambda^{2}}$ and one
$U(\varphi)=-M^{2}ch^{2}\mu^{'}\varphi$  at $\varphi^{2}=\frac{2
M^{2}ch^{2}\mu^{'}}{\lambda^{2}}$. It results to that
$L_{tot}(\varphi)_{Higgs}=L^{0}_{maximon}(\varphi) $ i.e.
\begin{equation}
L^{0}_{maximon}(\varphi)=\frac{1}{2}(\frac{\partial
\varphi}{\partial x_{\mu}})^{2}-\frac{1}{2}M^{2}\varphi^{2}.
\end{equation}

Now we shall consider a case when $ L_{int}(\varphi,\chi)=-
\frac{\lambda^{2}}{4} \varphi^{2}\chi^{2}$ . At $m\rightarrow im$
lagrangian (5) shall look like:
\begin{equation}
L_{tot}(\varphi,\chi)=\frac{1}{2}[(\frac{\partial
\varphi}{\partial
x_{\mu}})^{2}+M^{2}sh^{2}\mu^{'}\varphi^{2}-M^{2}(\chi-ch\mu^{'}\varphi)^{2}-
\frac{\lambda^{2}}{2}\varphi^{2}\chi^{2}],
\end{equation} where $Msh\mu^{'}=m$ . If we
differentiate (12) with respect to $\chi$, we find:
$$
\chi=\frac{M^{2}ch\mu^{'}\varphi}{M^{2}+\frac{\lambda^{2}}{2}\varphi^{2}}.
$$
Now (12) shall look like:
\begin{equation}
L_{tot}(\varphi)=\frac{1}{2}[(\frac{\partial \varphi}{\partial
x_{\mu}})^{2}+M^{2}sh^{2}\mu^{'}\varphi^{2}-\frac{\lambda^{2}}{2}\varphi^{2}(
\frac{\lambda^{2}\varphi^{2}}{2M^{2}}+1)\frac{M^{4}ch\mu^{'}\varphi^{2}}{(M^{2}+
\frac{\lambda^{2}\varphi^{2}}{2})^{2}}].
\end{equation}
This is one of Higgs generalizations on fundamental mass. From
(13) at $M\rightarrow\infty$  we shall receive the usual Higgs
lagrangian:
  \begin{equation}
\lim_{M\rightarrow\infty}
L_{tot}(\varphi)=\frac{1}{2}[(\frac{\partial \varphi}{\partial
x_{\mu}})^{2}+m^{2}\varphi^{2}-\frac{\lambda^{2}\varphi^{4}}{2}].
\end{equation}

$$
L_{tot}(\varphi)=\frac{1}{2}(\frac{\partial \varphi}{\partial
x_{\mu}})^{2}- M^{2}\varphi^{2},
$$ i.e.
$$
L_{tot}(\varphi)_{Higgs}=L^{0}_{maximon}(\varphi)
$$
     In case of spontaneous breaking of global symmetry $U(1)$ we
have:
 $$
L_{tot}(\varphi,\chi)=|\frac{\partial \varphi}{\partial
x}|^{2}-m^{2}|\varphi|^{2}-M^{2}|\chi-\cos\mu\varphi|^{2}
-\frac{\lambda^{2}}{2}|\chi|^{2}|\varphi|^{2}=
$$
\begin{equation}
=|\frac{\partial \varphi}{\partial
x}|^{2}-M^{2}|\varphi|^{2}-M^{2}|\chi|^{2}+M^{2}\cos\mu|\chi\overline{\varphi}+
\varphi\overline{\chi}|-\frac{\lambda^{2}}{2}|\chi|^{2}|\varphi|^{2}
\end{equation}
at $m\rightarrow im$  , then
\begin{equation}
L_{tot}(\varphi,\chi)=|\frac{\partial \varphi}{\partial
x}|^{2}-M^{2}|\varphi|^{2}-M^{2}|\chi|^{2}+M^{2}ch\mu^{'}|\chi\overline{\varphi}+
\varphi\overline{\chi}|-\frac{\lambda^{2}}{2}|\chi|^{2}|\varphi|^{2}.
\end{equation}
This lagrangian differs from (15) by its sign before $m^{2}$ , but
still invariant to group of global transformations:
\begin{equation} \begin{array}{ccc}
\varphi(x)\rightarrow\varphi(x)=\exp(igs)\varphi(x), \qquad \qquad
\varphi^{*}(x)\rightarrow\varphi^{*}(x)=\exp(-igs)\varphi^{*}(x)\\

\chi(x)\rightarrow\chi(x)=\exp(i g s)\chi(x), \qquad \qquad
\chi^{*}(x)\rightarrow\chi^{*}(x)=\exp(-i g s)\chi^{*}(x).
\end{array}
\end{equation}
Taking a derivative from (16) with respect to  $\chi$  and
$\overline{\chi}$, we shall find the equation of motion for $\chi$
and $\overline{\chi}$ accordingly:
$$-M^{2}\overline{\chi}+M^{2}ch\mu^{'}\overline{\varphi}-
\frac{\lambda^{2}}{2}|\varphi|^{2}\overline{\chi}=0 $$ and
$$-M^{2}\chi+M^{2}ch\mu^{'}\varphi-
\frac{\lambda^{2}}{2}|\varphi|^{2}\chi=0.
$$
 From these equations
we find:
$$
\overline{\chi}=\frac{ch\mu^{'}\overline{\varphi}}{1+\frac{\lambda^{2}}{2M^{2}}|\varphi|^{2}}
$$ and $$
\chi=\frac{ch\mu^{'}\varphi}{1+\frac{\lambda^{2}}{2M^{2}}|\varphi|^{2}}.
$$

Having substituted these values in (16), we shall find:
     \begin{equation}
L_{tot}(|\varphi|)=|\frac{\partial \varphi}{\partial
x}|^{2}-V(|\varphi|)
\end{equation}                                                                     (18)
where  $V(\varphi)$ is Higgs potential
$$
V(|\varphi|)=M^{2}|\varphi|^{2}-\frac{M^{2}ch^{2}\mu^{`}|\varphi|^{2}}{1+
\frac{\lambda^{2}}{2M^{2}}|\varphi|^{2}}.
$$
This potential has the minimum
$V_{min}(|\varphi|)=-\frac{2M^{4}}{\lambda^{2}}(ch\mu^{'}-1)^{2}$
at $|\varphi|^{2}=\frac{2M^{2}}{\lambda^{2}}(ch\mu^{'}-1)$. In a
flat limit $M\rightarrow\infty$  (18) will have a usual form. If
we shall write as  $
V_{New}(|\varphi|)=V(|\varphi|)-V_{min}(\varphi)$ then we have:
\begin{equation}
V_{New}(|\varphi|)=\frac{\lambda^{2}}{2}\frac{[|\varphi|^{2}-\frac{h^{2}}{2}]^{2}}{1+
\frac{\lambda^{2}}{2M^{2}}|\varphi|^{2}},
\end{equation}
where $\frac{h^{2}}{2}=\frac{2M^{2}}{\lambda^{2}}(ch\mu^{'}-1)$ ,
this quantity at $M\rightarrow\infty$  is equal
$\frac{m^{2}}{\lambda^{2}}$. At $M\rightarrow\infty$   (19) has a
usual form $$\lim_{M\rightarrow\infty}V_{New}(|\varphi|) =
\frac{\lambda^{2}}{2}[|\varphi|^{2}-\frac{m^{2}}{\lambda^{2}}]^{2}.$$
It is obvious, that function (19) has a minimum at
$|\varphi_{0}|^{2}=\frac{2M^{2}}{\lambda^{2}}(ch\mu^{'}-1)$. It is
always possible to choose as vacuum material value
$\varphi_{0}=\frac{\sqrt{2}M}{\lambda}\sqrt{ch\mu^{'}-1}.$ For
$L_{tot}(|\varphi|)$ we receive expression:
  \begin{equation}
L_{tot}(|\varphi|)=|\frac{\partial\varphi}{\partial x}|^{2}-
\frac{\lambda^{2}}{2}\frac{[|\varphi|^{2}-\frac{h^{2}}{2}]^{2}}{1+
\frac{\lambda^{2}}{2M^{2}}|\varphi|^{2}}.
\end{equation}
This lagrangian invariant to global gauge $U(1)$- transformation:
$\varphi(x)\rightarrow \varphi^{'}(x)=\varphi(x)e^{i\alpha}.$ The
system described by lagrangian (20), has spontaneous broken
symmetry $U(1).$ Now the point $ \varphi(x)=
\overline{\varphi^{*}}(x)=0$ does not corresponding with a minimum
of energy. There any point on a circle of radius $
R=\sqrt{2}\frac{M^{2}}{\lambda}\sqrt{ch\mu^{'}-1}, $  is agree
with a minimum of energy. We can choose as stable vacuum any
position, situated on a circle of radius $ R,$ that is all states
are equivalent because of change concerning transformation
$\varphi(x)\rightarrow \varphi^{'}(x)=\varphi(x)e^{i\alpha}.$ We
shall choose value of gauge phases $\alpha=0$ , uniform for all
the world, and we shall write down $ \varphi(x)$ in the form of
real and imaginary parts:
\begin{equation}
\varphi(x)=\frac{1}{\sqrt{2}} (h+\varphi_{1}(x)+i\varphi_{2}(x)),
\end{equation}
here $\varphi_{1}(x)$ and $\varphi_{2}(x)$ are two material
fields, describing excitation of system concerning vacuum
$\varphi(x)=\frac{h}{\sqrt{2}}.$ At transition to stable vacuum
$U(1)$ invariance is broken, as the phase of function $\varphi(x)$
is fixed.

In new variables for lagrangian (20) we have:
$$
L_{tot}(\varphi)\Rightarrow L_{tot}(\varphi_{1},\varphi_{2})=
$$
 \begin{equation}
=\frac{1}{2}[\frac{\partial\varphi_{1}(x)}{\partial x_{\mu}}]^{2}+
\frac{1}{2}[\frac{\partial\varphi_{2}(x)}{\partial x_{\mu}}]^{2}-
\frac{\lambda^{2}}{8}\frac{[\varphi_{1}^{2}+2h\varphi_{1}+\varphi_{2}^{2}]^{2}}{1+\frac{\lambda^{2}}{4
M^{2}}[(h +\varphi_{1})^{2}+\varphi_{2}]^{2}}
\end{equation}
As a result of spontaneous breaking of symmetry the goldstone
scalar massless particle $\varphi_{2}$  and the real scalar
particle  $\varphi_{1}$  with mass \begin{equation}
m_{1}=\frac{\lambda h}{[1+\frac{\lambda^{2}h^{2}}{4 M^{2}}]^{1/2}}
\end{equation} have appeared. At $M\rightarrow\infty$ we have $m_{1}=\sqrt{2}m. $

Author would like to thank sincerely  A.Di Giacomo, V. Gogohia,
M.Ausloos, R.Ibadov, M.Musakhanov and N.Takibaev  for useful
arguing of outcomes, constant attention and help with a spelling
of the given work.

\end{document}